# An Energy Efficient Wearable for Monitoring Elderly People Health

Maicon Much[1], Ayalon de Moraes Filho[1], Julio Sieg[1], Fabiano Hessel[1], Alfredo Cataldo Neto[2], César Marcon[1]

[1] Graduate Program in Computer Science at Pontifical Catholic University of Rio Grande do Sul (PPGCC/PUCRS)
[2] Graduate Program in Biomedical Gerontology at Pontifical Catholic University of Rio Grande do Sul (GERONBIO/PUCRS)

***Abstract*— The aged population is exposed to risk situations, like falls, sudden changes in vital signs, and faints. These situations become common at this life stage due to the body capacity decrease and illnesses increase. This scenario imposes attention on families and the national health system. Some studies propose to overcome this problem through elderly health monitoring systems based on smartphones, smartwatches, and other equipment that usually covers low efficient communication systems and low battery life. This paper proposes an efficient health monitoring system based on Sigfox IoT service and a watch-type wearable device for sensing elderly motion, location, and vital signs to predict risk situations designed to achieve at least 30-days of battery life. Our experimental results show that this architecture is of low cost, low power, efficient and feasible elderly health monitoring solution.**

## I. Introduction

Approximately 28-35% of people aged 65 or over have falls every year [1][2], increasing to 32-42% for those over 70 years of age [3], becoming fall the second leading cause of accidental or unintentional injury deaths worldwide [4]. Besides fall situations, older people are more suitable to suffer from a heart attack, low or high blood pressure [5], variations in skin temperature, depression, dementia, among other illnesses that can expose them to a fatal risk situation. Additionally, the number of older people living alone is continually increasing and associated with the advances in diagnosis and treatment of diseases; the life expectancy is rising and creating a global scenario of older people away from an assisted living.

Thanks to the new technology of microscopic devices, namely Micro-Electro-Mechanical Systems (MEMS), a high number of small-size light-weight wearable fall detectors have been developed over the last years. These devices are based on accelerometers, gyroscopes, and other sensors that intend to acquire information from users to detect any abnormal situation continuously.

Wearable energy consumption has a meaningful impact on its elderly acceptability because combining low battery capacity with a high-energy consumption forces a high frequency of recharging the device, an undesirable characteristic.

This work proposes a feasible and acceptable wearable for elderly monitoring falls and vital signs during an extended time without charging.

## II. Related Work

Nowadays, older people use smartphones to make a call or send emergency messages. However, they may not be able to use this communication in certain situations, requiring an automated risk detection system to contact the family or health system. Automatic fall detection systems based on wearables or smartphones are becoming a frequent research topic in the past decade, generating many publications in this area. Analyzing these publications, we observe that they are usually based on a Bluetooth communication system [6][7] that limits their capacity to the paired smartphone range or phone network [8] that is known as an inefficient communication system in terms of energy consumption.

The Internet of Things (IoT) enables the development of tiny devices with high communication capacity and low energy consumption. Wearable sensors [9] and video-based systems [10] are the most common techniques used to collect user data for detecting risk situations in older people, usually fall detection. Video-based systems are no longer applicable because of the monitoring limitation and privacy invasion [11]; still, wearable-based systems are becoming a popular research topic due to their non-invasive characteristics and promising results. Bhoi et al. [12] developed an IoT elderly fall detection system based on machine learning algorithms applied to accelerometer and gyroscope sensors using the Arduino platform to achieve valuable results. However, they employed *wifi* standard as the IoT communication connected to a local network for sending messages to the cloud. Queralta et al. [13] detailed the IoT characteristics of the LoRa [14] communication system as limited data rate and coverage area; that have similar characteristics with Sigfox, except in coverage area that in Sigfox is well explored. They used recurrent neural networks applied to motion sensors to detect falls locally, overcoming bandwidth limitations by sending fewer data through the network.

TABLE I lists important research in wearable monitoring solutions with a focus on low energy consumption. As reported, energy-saving is considered open research for wearables because they are highly dependent on the frequent recharging process. The emerging of IoT radio-frequency communication technology used in [15] and [16] brought numerous advances

TABLE I
ELDERLY MONITORING SYSTEMS BASED ON BATTERY POWERED WEARABLE

| Work | Communication technique | Battery capacity | Estimated battery lifetime |
|---|---|---|---|
| [15] | IoT[a] | 450 mAh | 48 hours |
| [16] | IoT | 500 mAh | 8 hours |
| [17] | Zigbee | 2000 mAh | 70 hours |
| [18] | Wifi/Bluetooth | 1260 mAh | 36 hours |
| This paper | Sigfox | 100 mAh | 720[b] hours |

a – Internet of Things communication that operates in unlicensed frequency spectra
b – Considering normal monitoring operation mode in power management, comparing with old technologies that present low performance in energy saving [17][18]. This paper introduces the LifeSenior system, a complete health monitoring system including specific care in energy consumption composed of a wearable with Sigfox communication system [19], the first global IoT network, sending emergency information directly to a cloud system, without connection with any smartphone or at-home network, achieving around of 30 days of battery lifetime.

## III. LifeSenior Wearable Architecture

LifeSenior is a global innovation because it is the first wearable IoT with native Sigfox communication embedded. It was developed to be a personal monitor for older people, pretty like an accessory and not comparable to a medical product. Figure 1 displays the LifeSenior wearable with its main components.

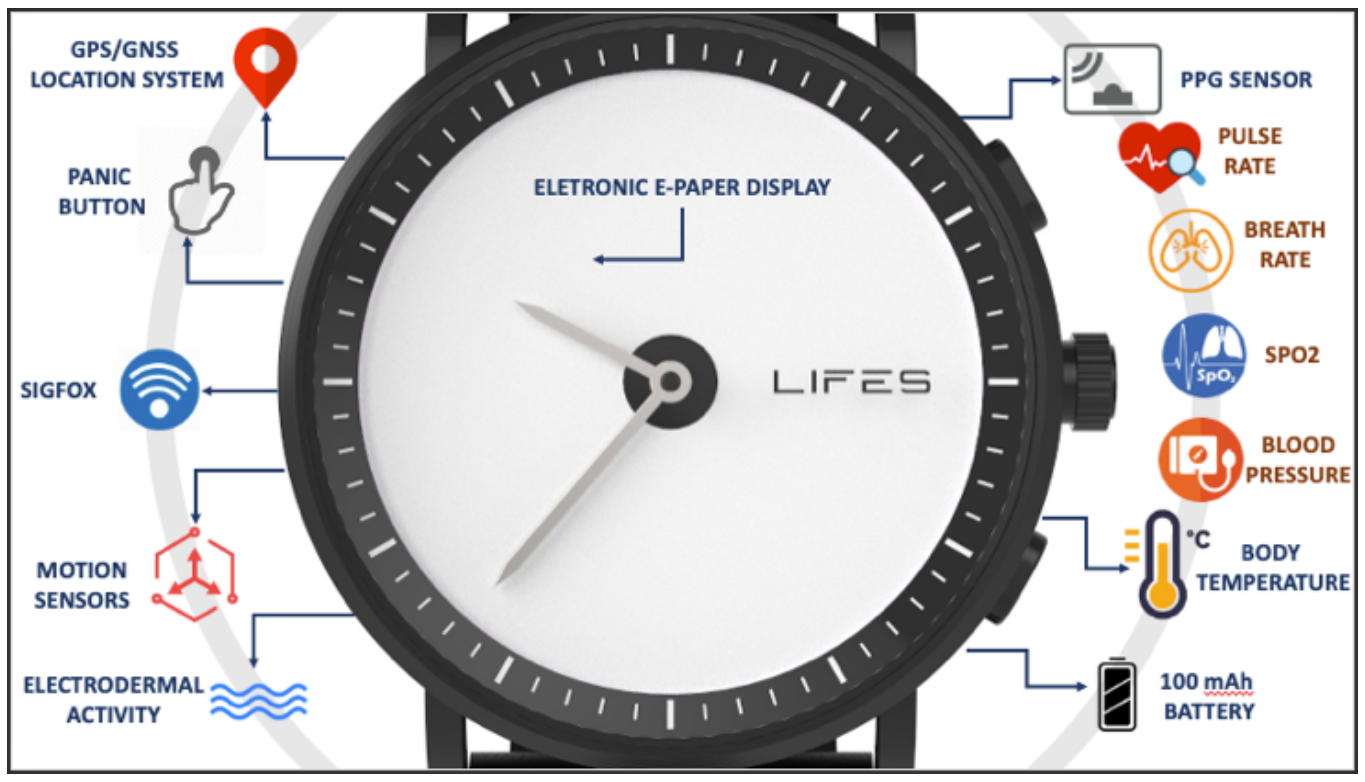

Figure 1. LifeSenior wearable device and its components. The location system is based on GPSS/GNS associated with Sigfox geolocation. When the Panic button is pressed, an emergency message is sent to the cloud. Motion sensors sense for fall problems, and the electrodermal (EDA) sensor monitors stress. a photoplethysmograph (PPG) and body temperature collect vital signs continuously. Segmented electronic e-paper display consumes energy only for a refresh. The battery powers the watch for 30-days in regular use.

Figure 2 shows the powered sensors, circuits, and components that need dedicated battery management and control to extend the battery lifetime. We chose the architecture components to meet the stringent energy-saving requirements and designed the management of these components to use activation and deactivation techniques when necessary.

Most parts of the circuits that compose LifeSenior are hot-start type, meaning they do not need to spend much time in start time. Joining this fact to the capacity to control each circuit individually and power them at the exact moment expected to acquire new information, we can adjust the energy consumption based on the specific moment in that each part is required.

The energy saver algorithm implemented in LifeSenior can be divided into hardware and software. Each command starts at the software layer at the application layer. The information sent by the application layer goes through the Hardware Abstract Layer (HAL) that interacts with each hardware part using a device driver and sends to the specific device in the circuit; the reverse direction also works. Figure 3 details the leading hardware and software layer and their interaction.

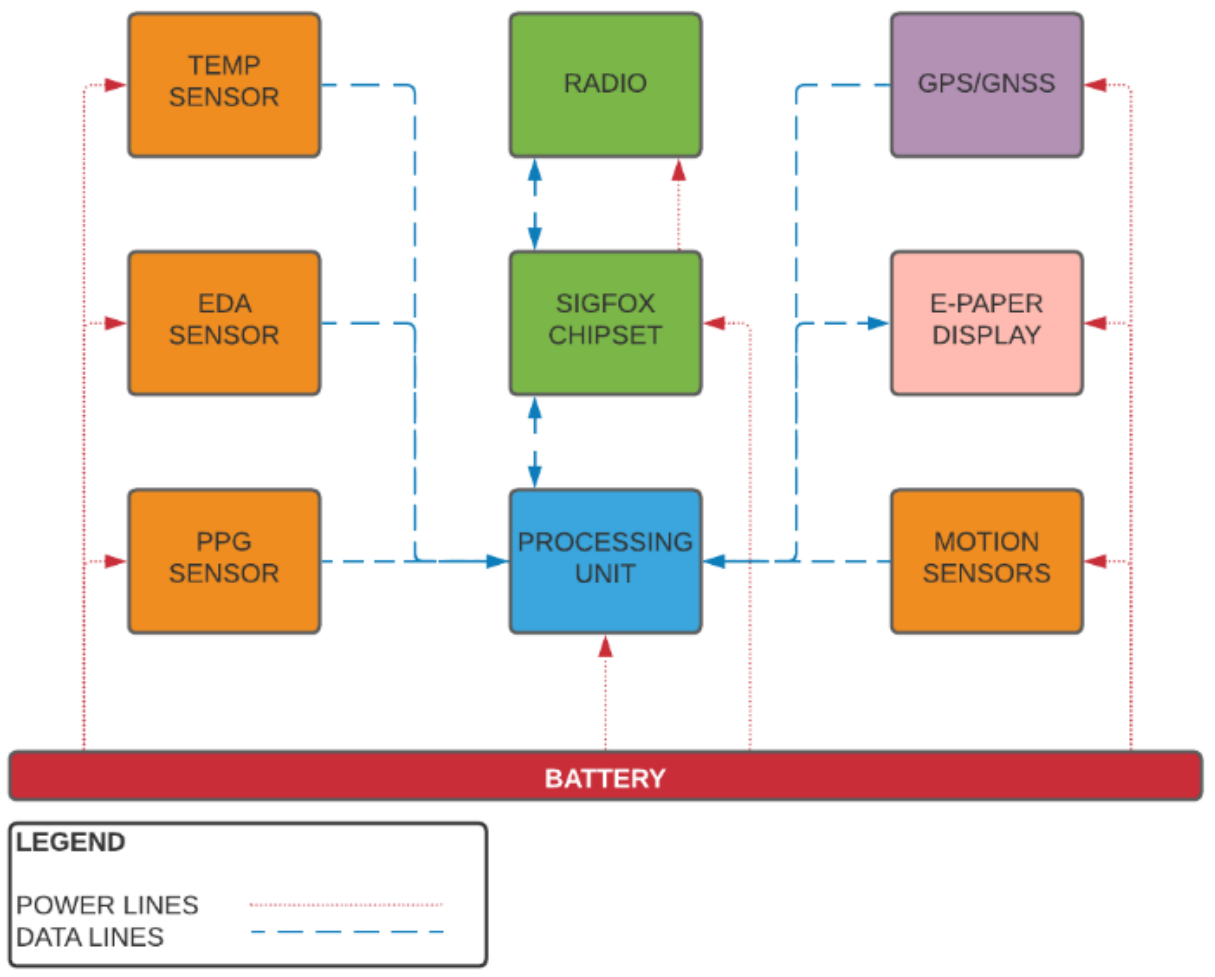

Figure 2. Wearable battery powered components.

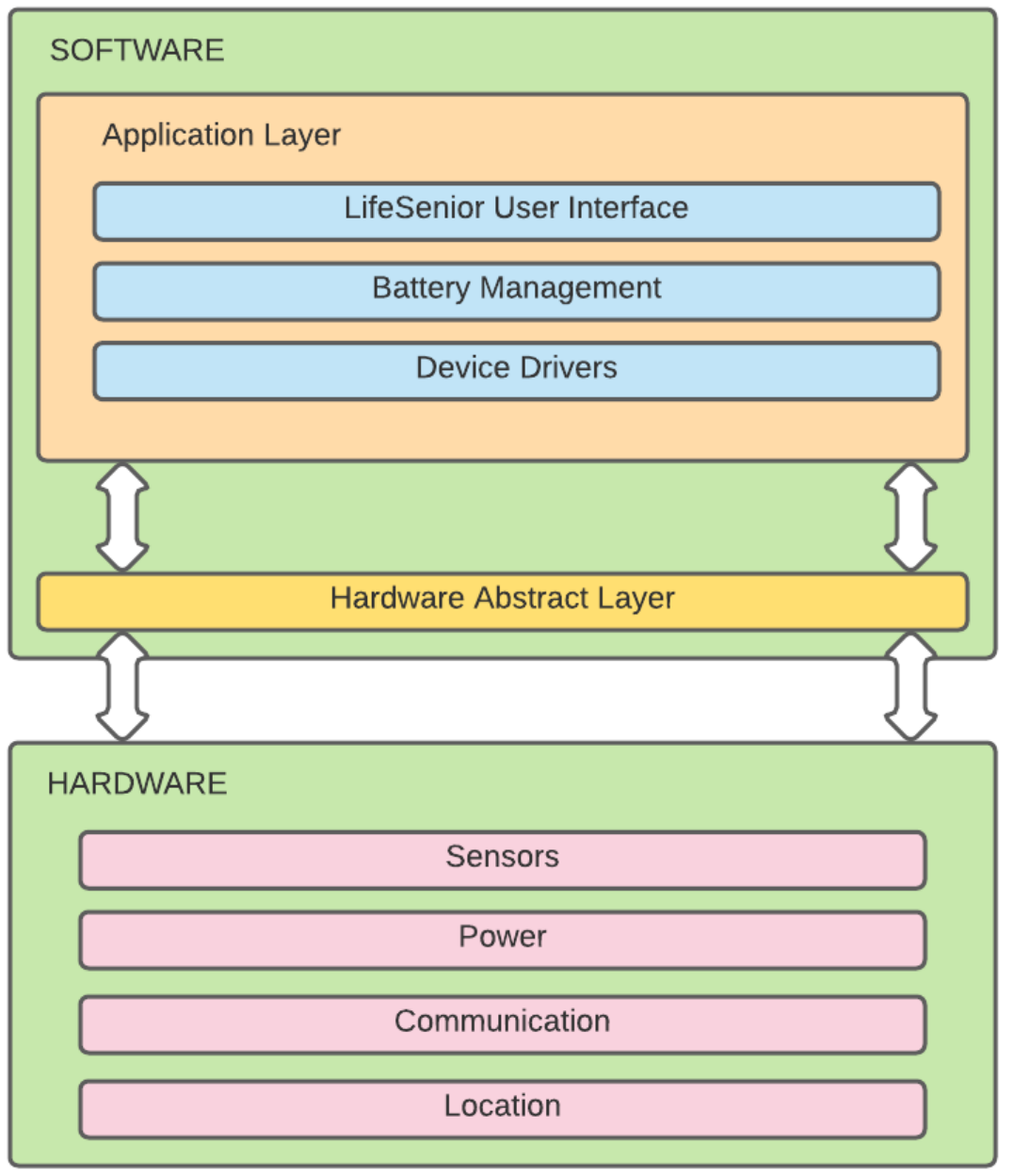

Figure 3. Detailed hardware and software layers and their communication systems.

### A. *Sigfox Communication System*

We choose Sigfox for implementing the wearable communication system since it is a low-power wireless network protocol aimed at IoT applications, gaining ground in several countries, like Brazil. According to WND, the company designing the Sigfox network in Brazil, there are more than 20 million devices connected to Sigfox in Brazil, and this number is increasing fast each year [20]. The main advantage of this protocol is operating with low energy consumption on unlicensed frequencies (e.g., 900 MHz in Brazil). Specifically, the low-power characteristic of Sigfox motivated our choice.

Sigfox offers software-based communication; i.e., all the data complexity is managed in the cloud rather than handled on the devices. This feature, specifically, helps the efficient energy consumption comparing to other communication systems. Sigfox consumes around 7 mA for less than a second, an incredible characteristic for the proposed battery management idea. Figure 4 shows the block diagram of the communication system, highlighting the enable circuit designed to turn on/off this system. Each message is prepared in the processing unit and sent to the communication block digitally. The Sigfox chipset converts this information to analog data and powers it with low energy to send the information to any radio base until 60 km of distance.

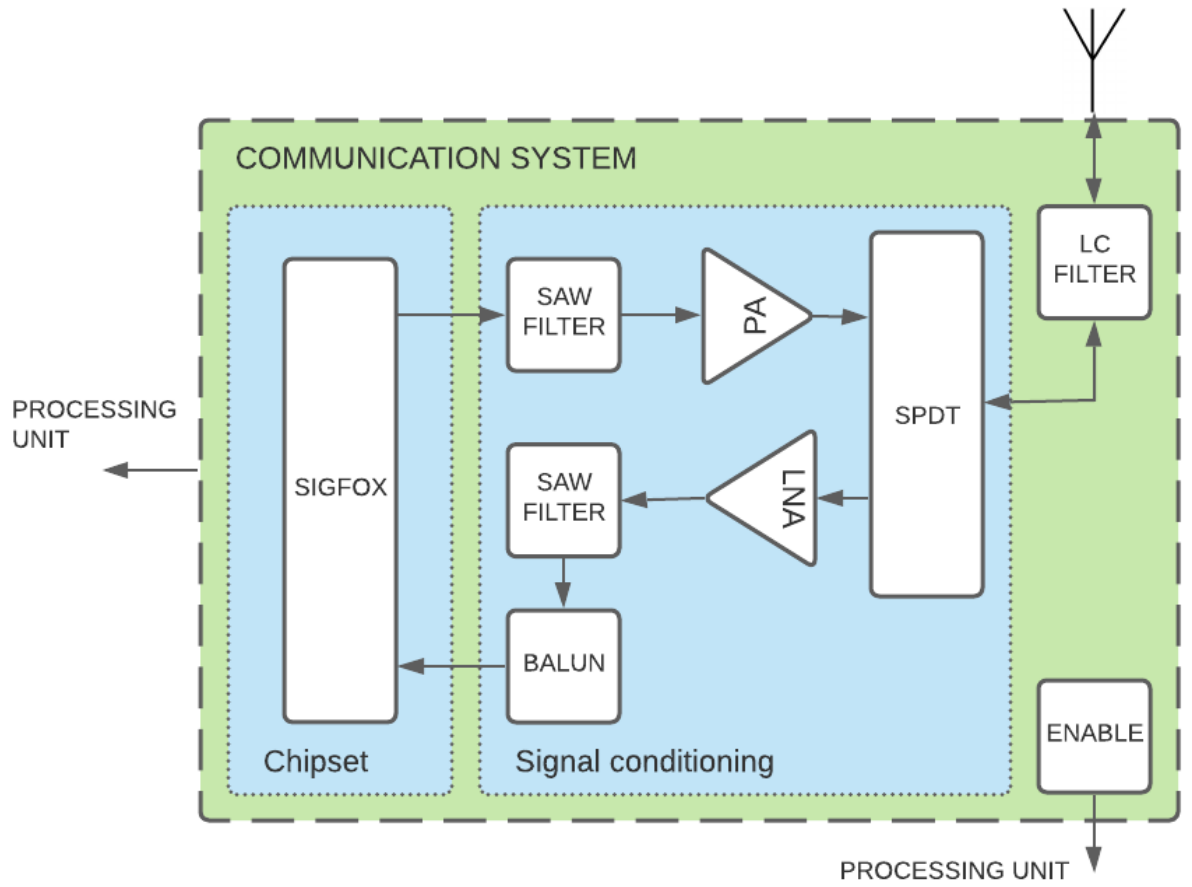

Figure 4. Block diagram of the LifeSenior communication system.

### B. Sigfox Circular Ring Antenna

Wearable antennas are essential components in various applications such as physiological sensors, intelligent wearable devices, and emergency rescue services. There has been increasing research interest in wrist-wear wireless communications, and the emergent technology has been deployed in various applications such as smartwatches.

A successful antenna design is essential to enhance smartwatch applications' communication quality because it can simultaneously fulfill several performance criteria. Specifically, the antenna must be compact and low profile.

Early attempts at developing smartwatch antennas have not been very successful because many structures do not fulfill the low-profile requirement because they need to attend to the vertical height restrictions imposed by the watchcase. The height is not the only restriction, but also the small diameter of the common watchcases limit space within the wearable device.

As the communication system of LifeSenior is on a Sigfox network that works in the 900 MHz ISM frequency band in Latin America (Sigfox RC2 for Brazil), the design of an efficient antenna is a big challenge. It can explain why most smart wearable devices today only have either a Bluetooth or a Wi-Fi antenna that needs less internal space than one for a 900 MHz frequency band with similar efficiencies.

Meanwhile, a low manufacturing cost and a low Specific Absorption Rate (SAR) are also highly desired since the human body is a lossy dielectric medium, the antenna performances must be robust against wrist tissue. Once the antenna is placed close to a human body, the radiation efficiency can be dramatically reduced.

To minimize the occupied space in the LifeSenior device, we proposed a circular ring antenna type made in copper and positioned immediately below the e-paper display with 33 mm ($0.1\lambda0$) in internal diameter, 2 mm ($0.006\lambda0$) thickness, and 3 mm ($0.009\lambda0$) high, as shown in Figure 5. The e-paper display was chosen with minimum metal parts and free of electronic components to reduce the electromagnetic interaction with the antenna, which certainly would degrade the communication performance.

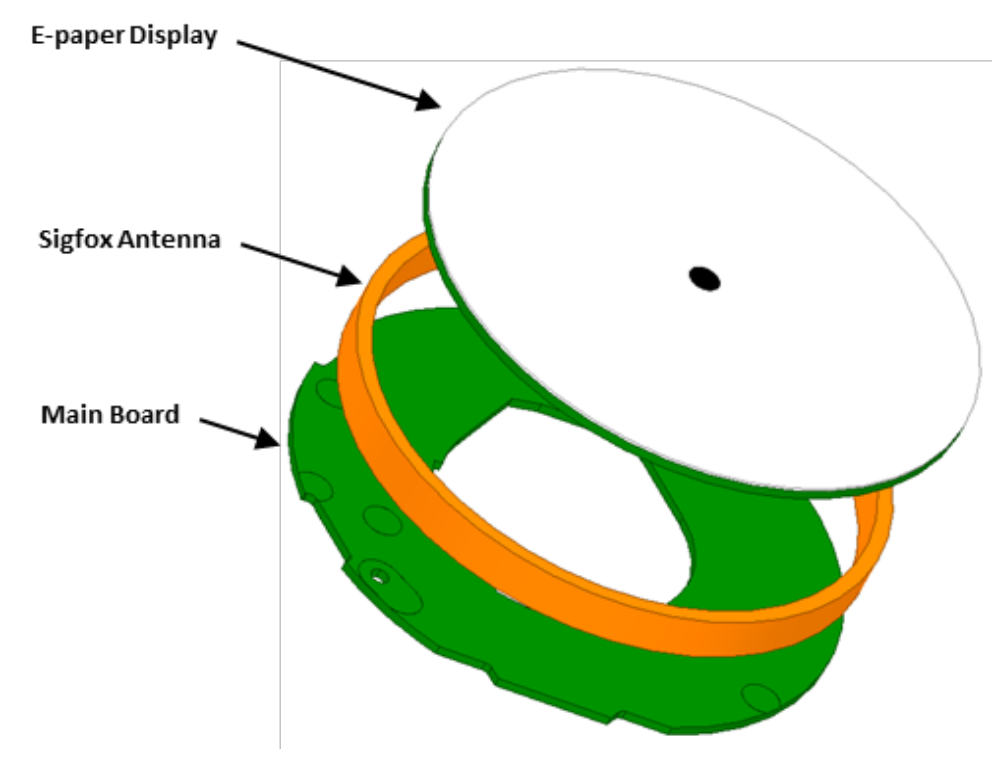

Figure 5. Sigfox antenna design.

### C. IoT Data Traffic

Sigfox protocol limits the packets sizes and the number of messages sent by day per device. Therefore, it is important to optimize the data transmission to meet this requirement and transmit as much information as possible in one shot to take advantage of the available energy. TABLE II details a single IoT transmission through the Sigfox network (12 bytes). The message is formed by:

- GPS [48 bits]: global geographic coordinates;
- EME [1 bit]: emergency button pressed;
- FAL [1 bit]: fall detected;
- BAT [5 bits]: battery level;
- SPO [7 bits]: oxygen saturation;

TABLE II
LIFESENIOR SIGFOX DETAILED PACKET FOR A SINGLE TRANSMISSION

| BYTE | BIT 7 | BIT 6 | BIT 5 | BIT 4 | BIT 3 | BIT 2 | BIT 1 | BIT 0 |
|---|---|---|---|---|---|---|---|---|
| 0 | GPS | GPS | GPS | GPS | GPS | GPS | GPS | GPS |
| 1 | GPS | GPS | GPS | GPS | GPS | GPS | GPS | GPS |
| 2 | GPS | GPS | GPS | GPS | GPS | GPS | GPS | GPS |
| 3 | GPS | GPS | GPS | GPS | GPS | GPS | GPS | GPS |
| 4 | GPS | GPS | GPS | GPS | GPS | GPS | GPS | GPS |
| 5 | GPS | GPS | GPS | GPS | GPS | GPS | GPS | GPS |
| 6 | EME | FAL | BAT | BAT | BAT | BAT | BAT | BAT |
| 7 | BAT | SPO | SPO | SPO | SPO | SPO | SPO | SPO |
| 8 | PR | PR | PR | PR | PR | PR | PR | PR |
| 9 | PR | BR | BR | BR | BR | BR | BR | BR |
| 10 | BR | EDA | EDA | EDA | EDA | EDA | EDA | EDA |
| 11 | EDA | EDA | - | - | - | - | - | - |

- PR [9 bits]: pulse rate;
- BR [8 bits]: breathing rate.
- EDA [9 bits]: electrodermal activity.

The communication architecture exposed was implemented in the wearable device, and some field tests are presented in the Experimental Results section.

### D. Location System

Worldwide, around 50 million people have dementia, and there are nearly 10 million new cases a year. Alzheimer's disease is the most common form of dementia and may contribute to 60-70% of the cases [21]. Most of these people stay alone at home for at least a part of the day and quickly get out and get lost. Trying to help this kind of people and their families, we employed in our wearable a GPS module (L96 module from Quectel company) to collect the user location. In the LifeSenior application, the responsible can determine an acceptable perimeter, and in case of violation, the responsible is notified and can follow the monitored person's location in real-time. Note that Sigfox also provides a location system based on radio base triangulation, but unfortunately, their precision is in kilometers order.

To verify our location system accuracy, we performed some tests that can be seen in the Experimental Results section.

### E. User Interface Display

We choose an e-paper display to show the wearable information. The most important advantage of this technology is that it consumes energy only during the refresh process, meaning that static information remains on the screen without consuming energy. We choose the ET011TT3 from the E-INK e-paper manufacturer. In most parts, this display is composed of plastic materials and associated with the possibility of turn-off the display without impairing view and meeting some radio frequency communication requirements.

### F. Sensors

The wearable device detailed in Figure 1 was designed to be similar to a traditional watch, improving older people's acceptance. However, the LifeSenior wearable is much more complex than a simple watch. Inside the wearable case, dedicated circuits acquire vital user signs through a photoplethysmograph (PPG), contact body temperature, motion, and ElectroDermal Activity (EDA) sensors.

PPG is used as an optical technique for detecting blood volume changes in the microvascular bed of tissue [22]. Our system uses the ADPD1080 PPG sensor from Analog Devices that processes periodic data collected from the monitored user. Based on the PPG signal, LifeSenior extracts motion tolerant heart rate frequency [23], enabling the application to draw a trend variance, essential to control the Heart Rate Variability (HRV). HRV is a recognized tool for the estimation of cardiac autonomic modulations [24]. Figure 6 displays de PPG sensor block diagram that includes an enable block to optimize the energy consumption. The light emitted by the TX block LED is reflected by the tissues and absorbed by the RX block photodiode. The relation between the light absorbed and overpassed is directly proportional to the physiological body response to oxygen transportation, enabling estimating vital signs, like heart rate.

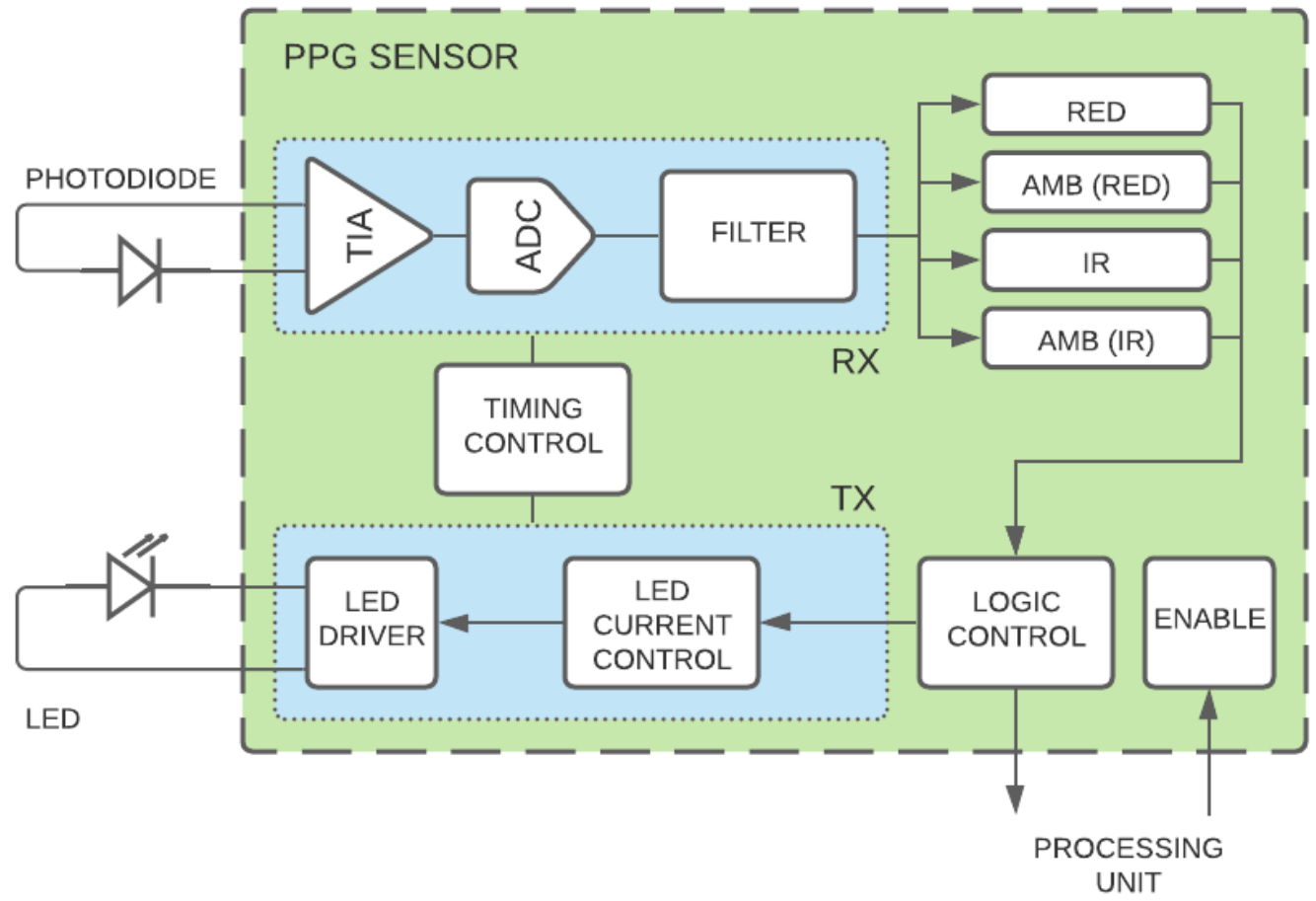

Figure 6. Photoplethysmograph (PPG) sensor block diagram.

Processing optic sensor data, the wearable device also estimates the user oxygen saturation ($SpO_2$) [25], filtering noise data to provide clean and trusted $SpO_2$ information. A decrease in $SpO_2$ value can indicate a reduction in oxygen circulation, which can predict respiratory problems. Besides pulse rate and $SpO_2$, LifeSenior has a way to estimate the user Breathing Rate (BR) [26], a crucial physiological parameter used in a range of clinical settings, including patient deterioration analysis.

The user temperature is also continuously collected, helping the health status algorithm to detect any abnormal scenario. The circuit responsible for this is detailed in Figure 7. LifeSenior has a thermal pad in contact with the pulsing skin that balances its temperature over the use, modifying the resistance measured by the analog to digital converter becoming possible to check the body temperature continuously.

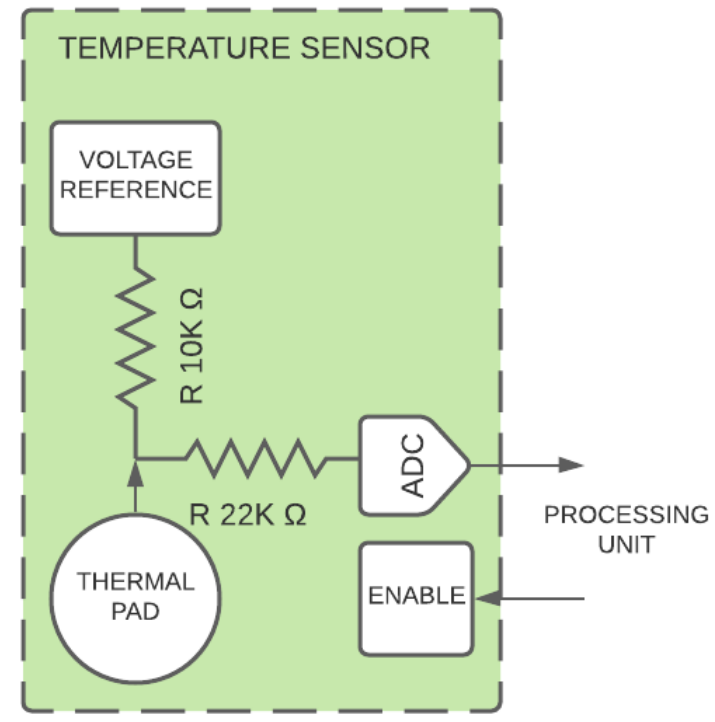

Figure 7. Temperature sensor.

Figure 8 details the EDA sensor that measures the skin's electrical activity through changes in the conductivity of the sweat glands in contact with the sensor. This change is directly proportional to the individual's emotional state; thus, together with other vital signs, enabling identifying nervousness, stress, or onset of depression [27]. The circuit is composed of an excitation block responsible for generating a slight current that passes through the skin by electrodes and is measured at the reception block. Emotional variations change the impedance of sweat glands that, consequently, changes the current that arrives at the reception channel, becoming possible to take an indirect measure of the human emotional moment.

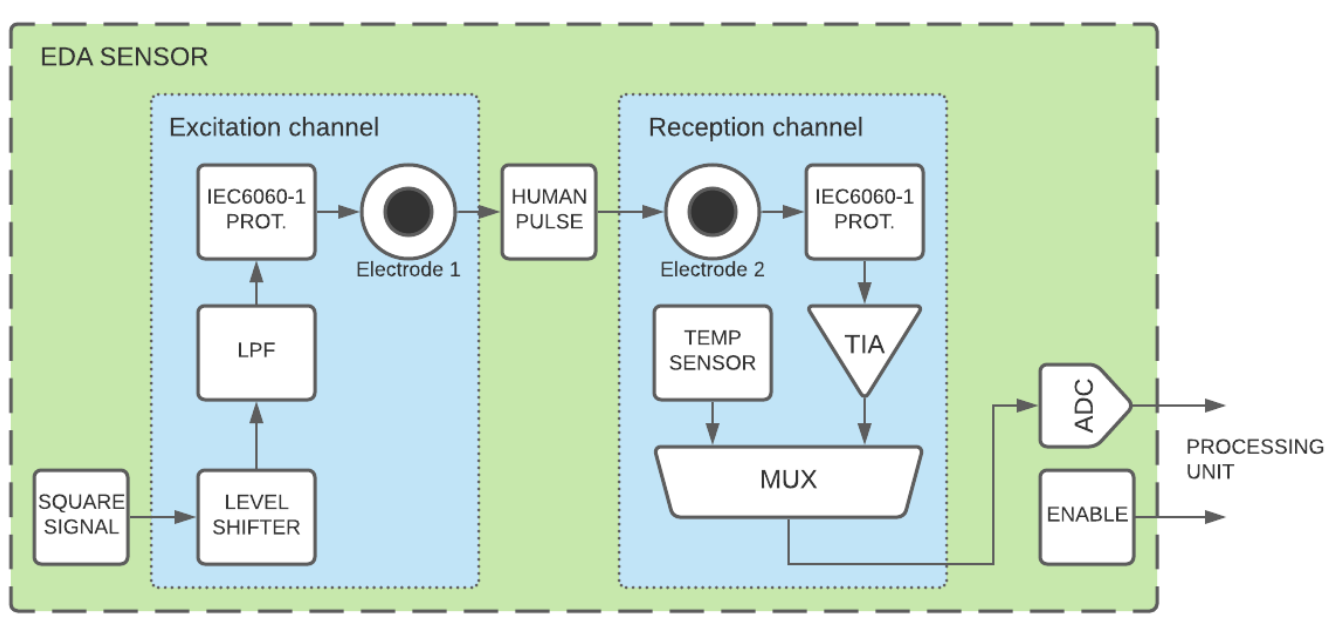

Figure 8. EDA sensor block diagram.

All physiologic data obtained by LifeSenior wearable is not collected in diagnostic character, but to monitor continuously the information checking for abnormal variations, rapid changes, or other situations that can indicate a health risk. Vital signs help LifeSenior to check the user-health condition and notify the family or health system in case of problems.

### G. *Battery Management*

We defined four user scenarios in which each peripheral can assume three operation modes to optimize the energy consumption of the LifeSenior wearable: (i) Standby or Low-power mode, (ii) Running soft processes, or (iii) Full processing. Depending on the specific need for hard, soft, or low processing, each circuit can be triggered, as we can see in the example shown in Figure 9. We estimate the total energy consumption for each user scenario during a day by combining each circuit's operation time with the necessary energy to complete each operation. Therefore, it is possible to check the goal of 30 days far from recharging is achieved.

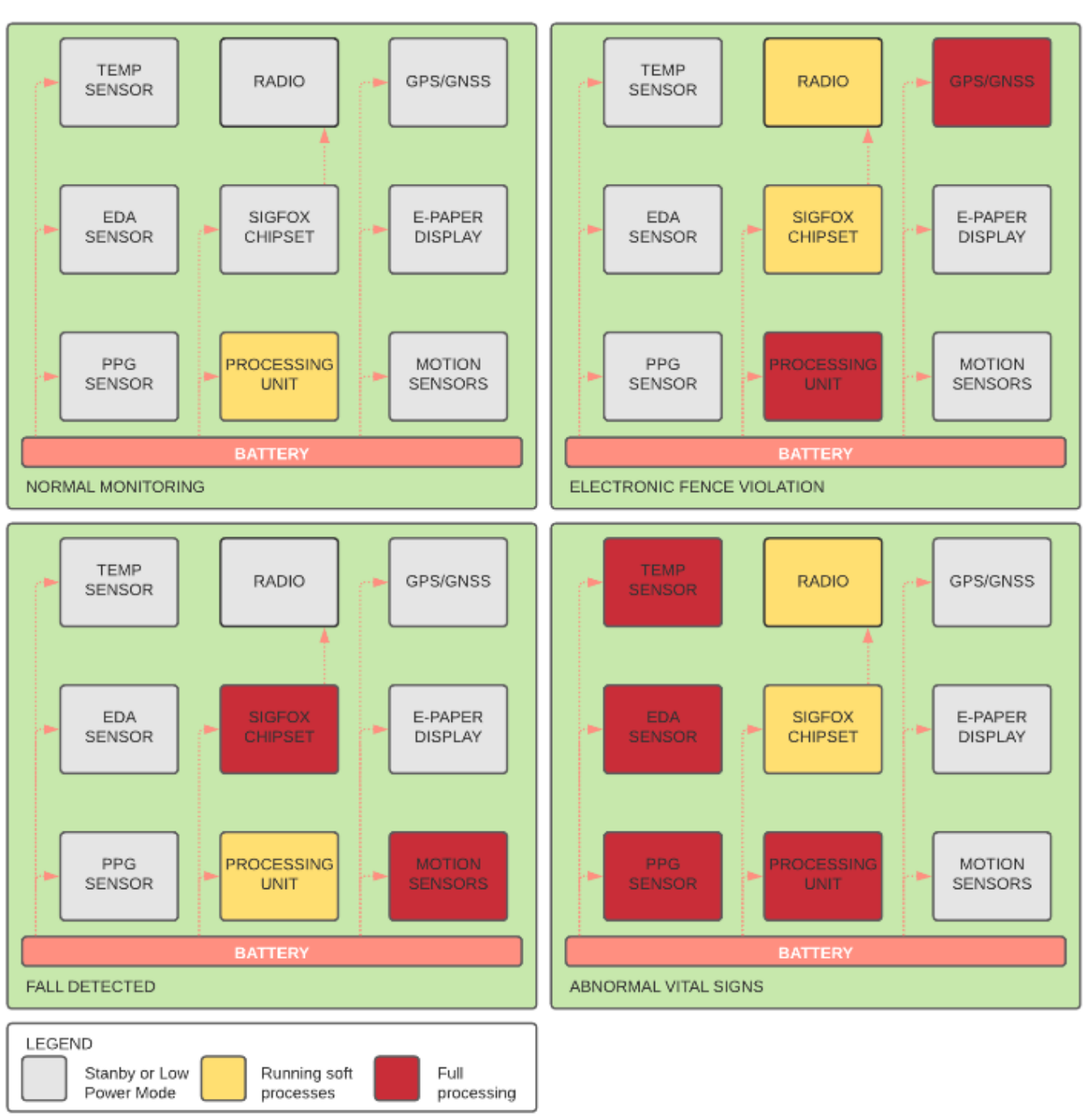

Figure 9. A) **Normal Monitoring**: Only processing unit is running soft processes; B) **Electronic fence violation**: in this scenario, GPS is running full processing mode because of the tracking mode activation; C) **Fall detected**: enable the motion sensors full operation; and D) **Abnormal vital signs**: enabling the vital signs sensors full mode.

TABLE III shows the four possible scenarios illustrated in Figure 9 of energy consumption of the LifeSenior wearable circuits. As expected, the normal monitoring energy consumption estimated for 30 days is less than the battery capacity (100 mAh), showing that it is possible to avoid often battery recharges. Nevertheless, the battery lifetime is reduced in critical scenarios, like electronic fence violation that requires turning on the GPS tracking mode.

## IV. Experimental Results

We conducted tests in relevant IoT scenarios for evaluating the communication and geolocation systems, enabling us to check the LifeSenior performance. We used a prototype that simulated all circuits and components described in this article in a standard case to perform these tests.

TABLE III
LifeSenior Estimated Energy Consumption for User Scenarios

| Mode | Current [mA] | Time [h] | Event [mAh] | Events per day | Day cons. [mAh] | 30-days [mAh] |
|---|---|---|---|---|---|---|
| | | | E-paper display | | | |
| A | 7 | 0.00027 | 0.0019 | 24 | 0.0046 | 1.4 |
| B | 7 | 0.00027 | 0.0019 | 24 | 0.0046 | 1.4 |
| C | 7 | 0.00027 | 0.0019 | 24 | 0.0046 | 1.4 |
| D | 7 | 0.00027 | 0.0019 | 24 | 0.0046 | 1.4 |
| | | | SigFox | | | |
| A | 10 | 0.00027 | 0.0027 | 12 | 0.033 | 1.00 |
| B | 10 | 0.00027 | 0.0027 | 30 | 0.046 | 2.50 |
| C | 10 | 0.00027 | 0.0027 | 40 | 0.111 | 3.33 |
| D | 10 | 0.00027 | 0.0027 | 30 | 0.046 | 2.50 |
| | | | GPS | | | |
| A | 0.5 | 0.0027 | 0.00135 | 24 | 0.0324 | 0.972 |
| B | 20.0 | 0.0800 | 1.60000 | 1 | 1.6000 | 48.000 |
| C | 0.5 | 0.0027 | 0.00135 | 24 | 0.0324 | 0.972 |
| D | 0.5 | 0.0027 | 0.00135 | 24 | 0.0324 | 0.972 |
| | | | PPG | | | |
| A | 4.5 | 0.01 | 0.037 | 24 | 0.899 | 26.98 |
| B | 4.5 | 0.01 | 0.037 | 24 | 0.899 | 26.98 |
| C | 4.5 | 0.01 | 0.037 | 24 | 0.899 | 26.98 |
| D | 4.5 | 0.01 | 0.037 | 48 | 1.790 | 53.97 |
| | | | Motion | | | |
| A | 0.017 | 1 | 0.017 | 24 | 0.408 | 12.24 |
| B | 0.017 | 1 | 0.017 | 24 | 0.408 | 12.24 |
| C | 0.550 | 1 | 0.550 | 3 | 1.650 | 61.50 |
| D | 0.017 | 1 | 0.017 | 24 | 0.408 | 12.24 |
| | | | Temperature | | | |
| A | 0.02 | 1 | 0.02 | 24 | 0.48 | 14.4 |
| B | 0.02 | 1 | 0.02 | 24 | 0.48 | 14.4 |
| C | 0.02 | 1 | 0.02 | 24 | 0.48 | 14.4 |
| D | 0.50 | 1 | 0.50 | 3 | 1.50 | 45.0 |
| | | | Processing Unit | | | |
| A | 0.00042 | 1 | 0.00042 | 24 | 0.01008 | 0.3024 |
| B | 2.24000 | 1 | 2.24000 | 1 | 2.24000 | 67.2000 |
| C | 0.00042 | 1 | 0.00042 | 24 | 0.01008 | 0.3024 |
| D | 2.24000 | 1 | 2.24000 | 1 | 2.24000 | 67.2000 |
| | | | EDA | | | |
| A | 8.2 | 0.0027 | 0.0227 | 24 | 0.54 | 16.350 |
| B | 8.2 | 0.0027 | 0.0227 | 24 | 0.54 | 16.350 |
| C | 8.2 | 0.0027 | 0.0227 | 24 | 0.54 | 16.350 |
| D | 8.2 | 0.0027 | 0.0227 | 48 | 1.09 | 32.708 |

A – Normal Monitoring energy consumption in 30 days: **73,65** mAh
B – Electronic Fence Violation energy consumption in 30 days: **189,08** mAh
C – Fall Detected energy consumption in 30 days: **125,25** mAh
D – Abnormal vital signs energy consumption in 30 days: **215,99** mAh

TABLE IV
RING ANTENNA CHARACTERISTICS

| Parameter | Value |
|---|---|
| Frequency band (MHz) | 900 ~ 908 |
| Voltage standing wave ratio (VSWR) | < 1.5 |
| Radiation efficiency (%) | 68% |
| Radiated power (dBm) @ 902.2 MHz | 20.2 |

### A. *Ring Antenna Performance*

The ring antenna was designed using a software tool based on a Method of Moments, a computational electromagnetic method. A usual impedance matching circuit was also designed to obtain the maximum power transfer between the antenna and Sigfox modem. Some meaningful results are presented on TABLE IV. It is worth note that despite the smartwatch's tiny dimensions, the antenna delivers a good radiation efficiency for the 900 MHz frequency band.

The ring antenna was implemented on the LifeSenior device, and laboratory tests performed to measure the radiated power showed promising results, as shown in TABLE V. Section IV.B describe some field tests for confirming the reliable design of the proposed antenna.

### B. *Field Tests with the Sigfox Communication System*

Sigfox network provides gateways scattered around the world to receive the messages sent by devices like LifeSenior. Therefore, to communicate with the cloud system, the device must be in a coverage area. We conduct most of the experimental tests in the coverage area of Porto Alegre city, Brazil, which presents an excellent Sigfox coverage, as one can see in the blue color detailed in Figure 10.

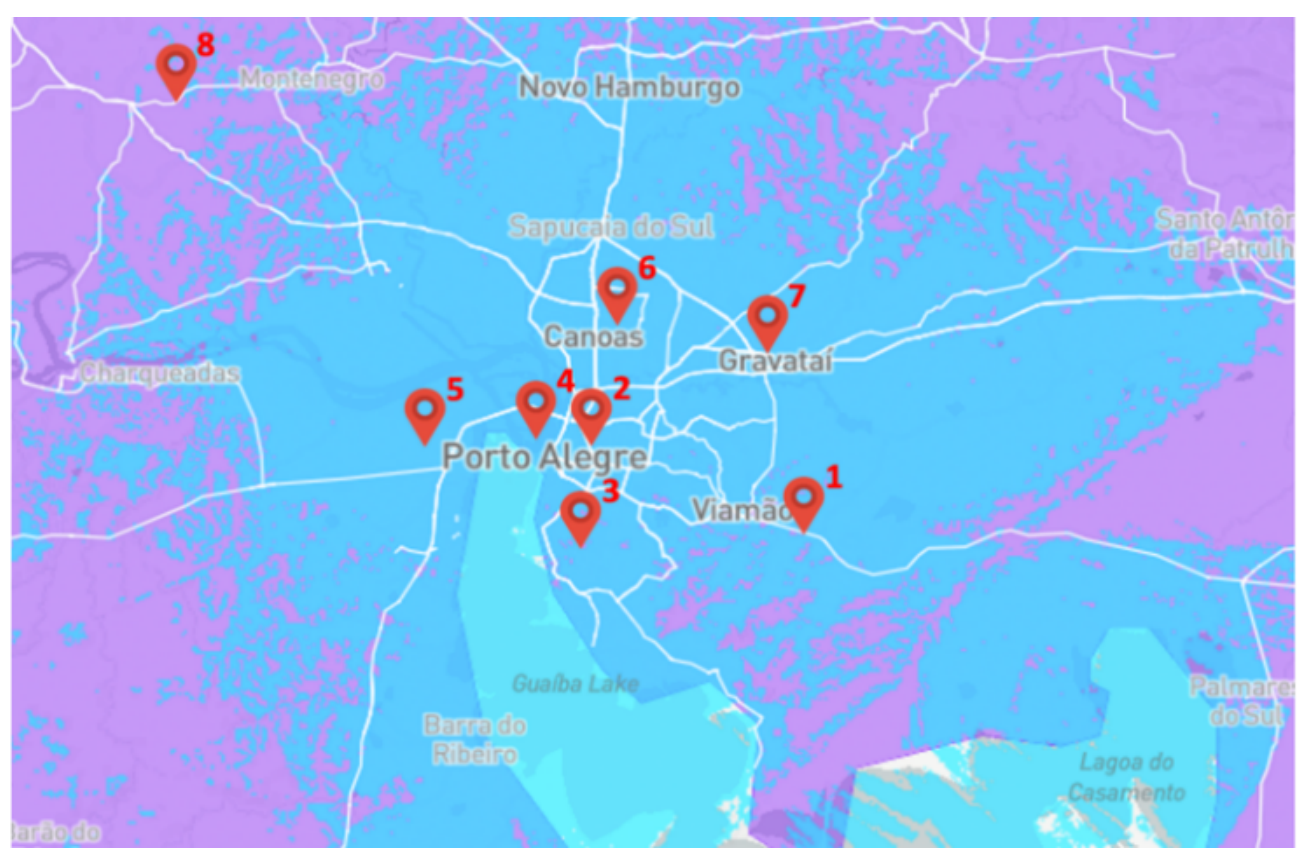

Figure 10. Sigfox coverage (in blue color) around of Porto Alegre city; tested points under the coverage area in red (GPS coordinates are in TABLE V).

The experiment consisted of going to a certain point of the map detailed on TABLE V, sending a message to the LifeSenior cloud network using our test device, and verifying if our cloud system received the message. TABLE V shows that all points under the coverage area were corrected received in our cloud system. Only messages sent outside the coverage area were not received, as expected.

TABLE V
SIGFOX NETWORK COVERAGE EXPERIMENTAL RESULTS

| Map point | GPS coordinate | Sigfox coverage | Message received |
|---|---|---|---|
| 1 | *-30.080911, -51.101350* | Yes | Yes |
| 2 | *-30.0309190, -51.205593* | Yes | Yes |
| 3 | *-30.122585, -51.212238* | Yes | Yes |
| 4 | *-29.996451, -51.214362* | Yes | Yes |
| 5 | *-30.000060, -51.318724* | Yes | Yes |
| 6 | *-29.909980, -51.175445* | Yes | Yes |
| 7 | *-29.944009, -50.984294* | Yes | Yes |
| 8 | *-29.686816, -51.493111* | No | No |

### C. *Location System*

We designed a mixed path, including building indoor location, underground point, and direct sky view, detailed in Figure 11 to test the wearable location accuracy in tracking mode. The test consisting of continuously getting coordinates from the GPS module shows that the coordinates received in indoor location and direct sky view followed the correct path (green and yellow points). However, some error was present when the underground point is crossed (red points).

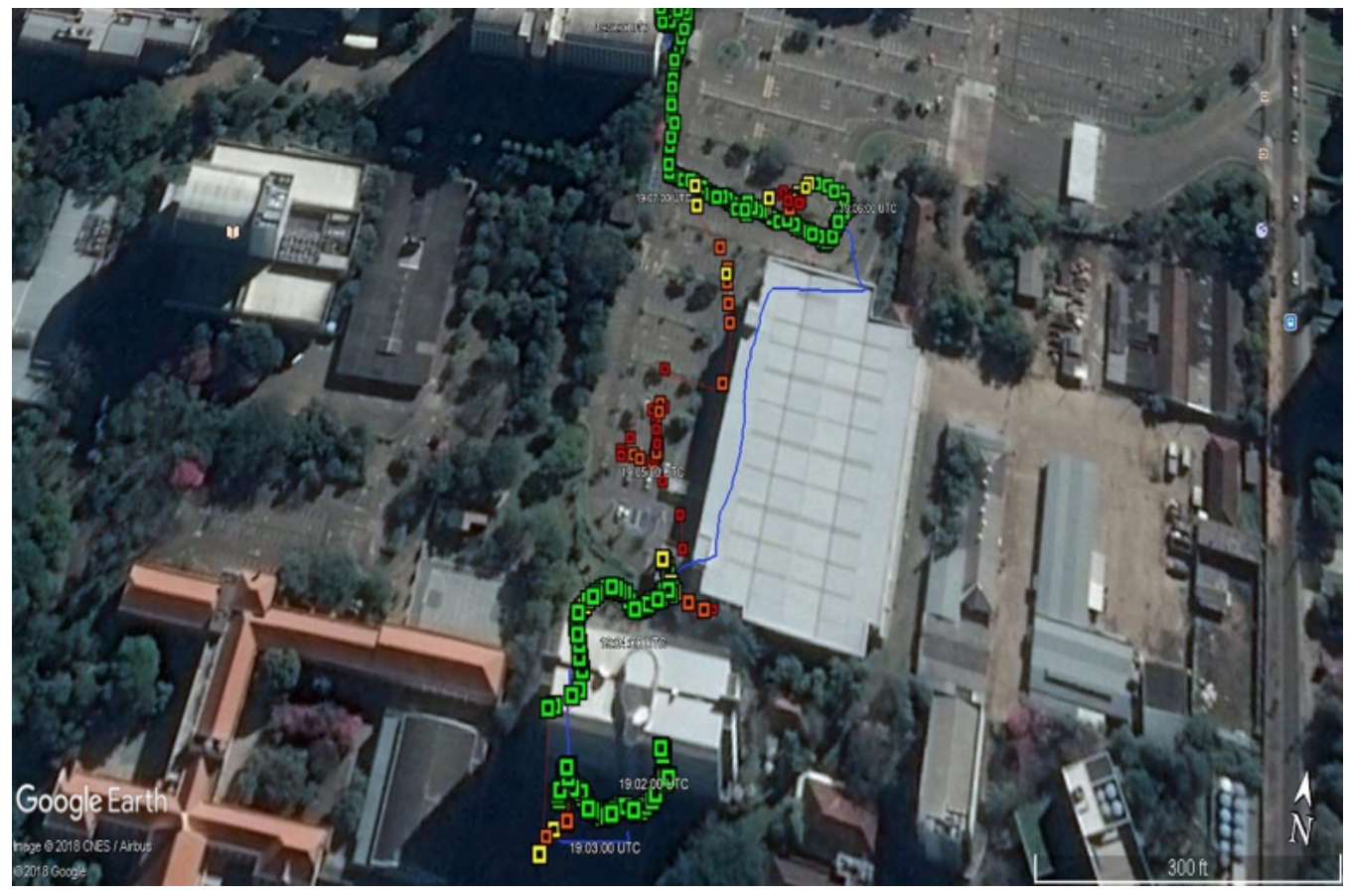

Figure 11. Performance of the location system test. Red points are positions with errors associated with the indoor location.

## V. CONCLUSION

We have proposed a system architecture for elderly health monitoring focusing on saving battery energy to achieve at least 30 days of monitoring without recharging the device. We have shown that the proposed system is feasible and can be used on a large scale, improving the quality of older people's lives.

We expect to develop new features in the device for future work, incrementing the number of sensors and studying other techniques to save more energy.

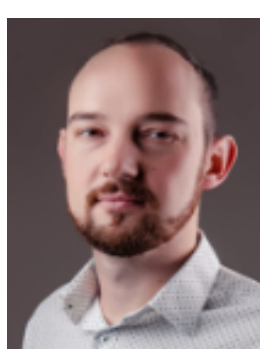

**Maicon Diogo Much** is Ph.D. student in computer science at Pontifical Catholic University of Rio Grande do Sul (PUCRS) researching technologies to improve the quality of elderly people life. He is M.Sc. and graduate in electrical engineering at PUCRS with focus in biomedical engineering. His research is focused in medical devices and vital signs monitoring, working in development process of medical products like patient monitors, defibrillators and health wearables.

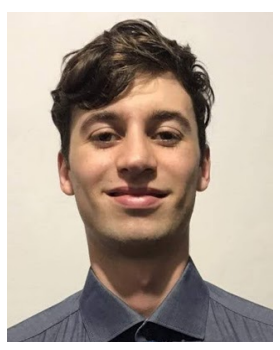

**Ayalon Angelo de Moraes Filho** received the B.E. degree in electrical engineering from Polytechnic School of Pontifical Catholic University of Rio Grande do Sul (PUCRS), Porto Alegre, Brazil, in 2020 and is currently studying M.S in computer science from Graduate Program in Computer Science (PPGCC) of PUCRS. His research interests are in the areas of embedded systems in the medical field. He is currently researching an algorithm to infer the respiration rate through the signal coming from the photoplethysmography sensor.

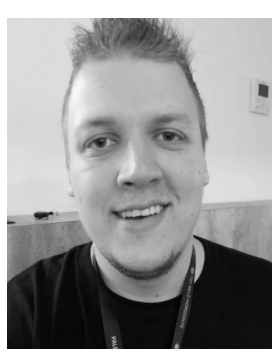

**Julio Alexander Sieg** received the B.S. degree in computer science from Passo Fundo University (UPF), Passo Fundo, Brazil, in 2017 and is currently studying M.S. in computer science from Graduate Program in Computer Science (PPGCC) of Pontifical Catholic University of Rio Grande do Sul, Porto Alegre, Brazil (PUCRS). He is currently researching daily living activities detection using a minimum quantity of wearable sensors.

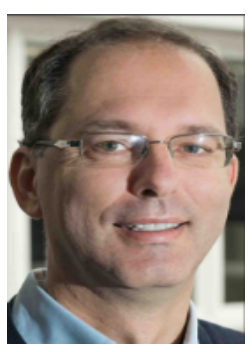

**Fabiano Passuelo Hessel** is Full Professor of Computer Science at PUCRS (Brazil) - School of Technology, Research Productivity Scholarship from CNPq. He was Advisor in the Office of the PUCRS Vice-President for Innovation, Research and Development. He received his Ph.D. in Computer Science from Joseph-Fourier University (now Université Grenoble Alpes), France (2000). He is the Coordinator of the Smart City Research and Innovation Center at PUCRS, and Elected Coordinator of the Advisory Committee on Mathematics, Statistics and Computer Science of the Foundation for Research Support of the State of Rio Grande do Sul. Professor Hessel is member of the IEEE. He was the Associate Editor of the ACM Transaction on Embedded Computer Systems, General Chair and/or Program Chair of several conferences and South America Representative of ISQED (International Symposium on Quality Electronic Design). He had several publications in prestigious conferences and journals, book chapters and books.

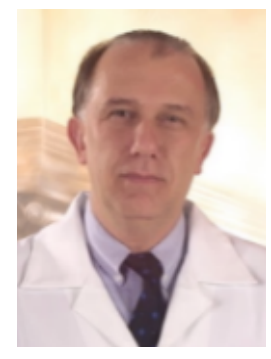

**Alfredo Cataldo Neto** is a Full Professor on the Biomedical Postgraduate Program in Gerontology at the Medical School of Pontifical Catholic University of Rio Grande do Sul (PUCRS), Brazil, since 1991. Doctor (Federal University of Rio Grande do Sul), specialist in Psychiatry (PUCRS, Brazilian Association of Psychiatry and World Psychiatry Association), Psychoanalyst (Psychoanalytic Society of Porto Alegre and International Psychoanalytic Association). He received his Ph.D. in Medicine from PUCRS in 1998. Coordinates the Aging and Mental Health Research Group (GPESM). He has 100 articles published in magazines, 108 book chapters, 64 completed doctoral, master's, specialization and scientific initiation orientations, 3 books and 40 scientific research titles and awards.

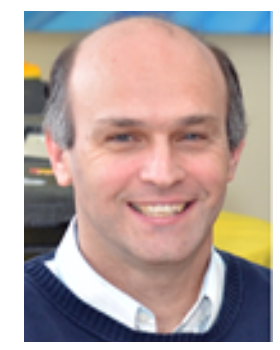

**César Marcon** (SM'19) is a Professor at the Polytechnic School of Pontifical Catholic University of Rio Grande do Sul (PUCRS), Brazil, since 1995. He received his Ph.D. in Computer Science from Federal University of Rio Grande do Sul, Brazil, in 2005. Professor Marcon is Senior Member of the Institute of Electrical and Electronics Engineers (IEEE) and of the Brazilian Computer Society (SBC). He is a Brazilian distinguished researcher with a CNPq PQ-2 grant. He is advisor of M.Sc. and Ph.D. graduate students at Graduate Program in Computer Science (PPGCC) of PUCRS. He has more than 150 papers published in prestigious journals and conference proceedings. Since 2005, prof. Marcon coordinated several research projects in areas of telecom and healthcare.